\documentclass[conference]{IEEEtran}
\IEEEoverridecommandlockouts
\usepackage{cite}
\usepackage{amsmath,amssymb,amsfonts}
\usepackage{algorithmic}
\usepackage{graphicx}
\usepackage{subcaption}
\usepackage{textcomp}
\usepackage{xcolor}
\usepackage{booktabs}
\usepackage{multirow}
\usepackage{url}
\usepackage{tabularx}
\usepackage{array}
\usepackage{makecell}

\def\BibTeX{{\rm B\kern-.05em{\sc i\kern-.025em b}\kern-.08em
    T\kern-.1667em\lower.7ex\hbox{E}\kern-.125emX}}

\begin{document}

\title{SoK: The Evolution of Maximal Extractable Value, From Miners to Cross-Chain}

\author{
\IEEEauthorblockN{Davide Mancino}
\IEEEauthorblockA{
\textit{University of Milano-Bicocca}\\
Milano, Italy\\
\texttt{davide.mancino@unimib.it}
}
\and
\IEEEauthorblockN{Hasret Ozan Sevim}
\IEEEauthorblockA{
\textit{University of Camerino}\\
\textit{Catholic University of Sacred Heart}\\
Camerino \& Milano, Italy\\
\texttt{hasretozan.sevim@unicam.it}
}
}

\maketitle

\begin{abstract}
This Systematization of Knowledge (SoK) provides a comprehensive historical analysis of Maximal Extractable Value (MEV) in blockchain systems, tracing its conceptual evolution through three distinct eras. We organize the fragmented literature on MEV into a unified chronological framework, beginning with Era~I (August 2014 -- August 2020), which introduced Miner Extractable Value from pmcgoohan's seminal Reddit warning through the ``Dark Forest'' recognition, covering Proof-of-Work systems with public mempools and Priority Gas Auctions. Era~II (August 2020 -- April 2024) marks the generalization to Maximal Extractable Value, encompassing formal taxonomies, Realized Extractable Value, Proposer-Builder Separation, the Ethereum Merge, MEV-Boost, and the integration of non-atomic and CEX-DEX arbitrage. Era~III (April 2024, present) addresses the frontier of Cross-Chain MEV, beginning with early studies on Layer-2 ecosystems, where value extraction spans multiple blockchains, rollups, bridges, and sequencers. We present a conceptual taxonomy distinguishing potential from realized extractable value, and single-domain from cross-domain phenomena. Our systematization identifies mitigations that emerged in response to each era, highlights measurement challenges, and proposes a research agenda for standardized metrics, detection benchmarks, and cross-chain infrastructure design.
\end{abstract}


\section{Introduction}
\label{sec:intro}

Maximal Extractable Value (MEV) has emerged as one of the most significant phenomena affecting the security, fairness, and economic dynamics of blockchain-based systems. Originally termed ``Miner Extractable Value'' in the context of Proof-of-Work (PoW) consensus, MEV refers to the profit that can be extracted by entities controlling transaction ordering within a block. As blockchain ecosystems have evolved, the idea of MEV and how it is measured has also changed, moving from an early concern about keeping the network stable to a multi-billion dollar industry with specialized players, advanced infrastructure, and even cross-chain strategies.

\begin{figure}[!ht]
\centering
\includegraphics[width=0.98\columnwidth]{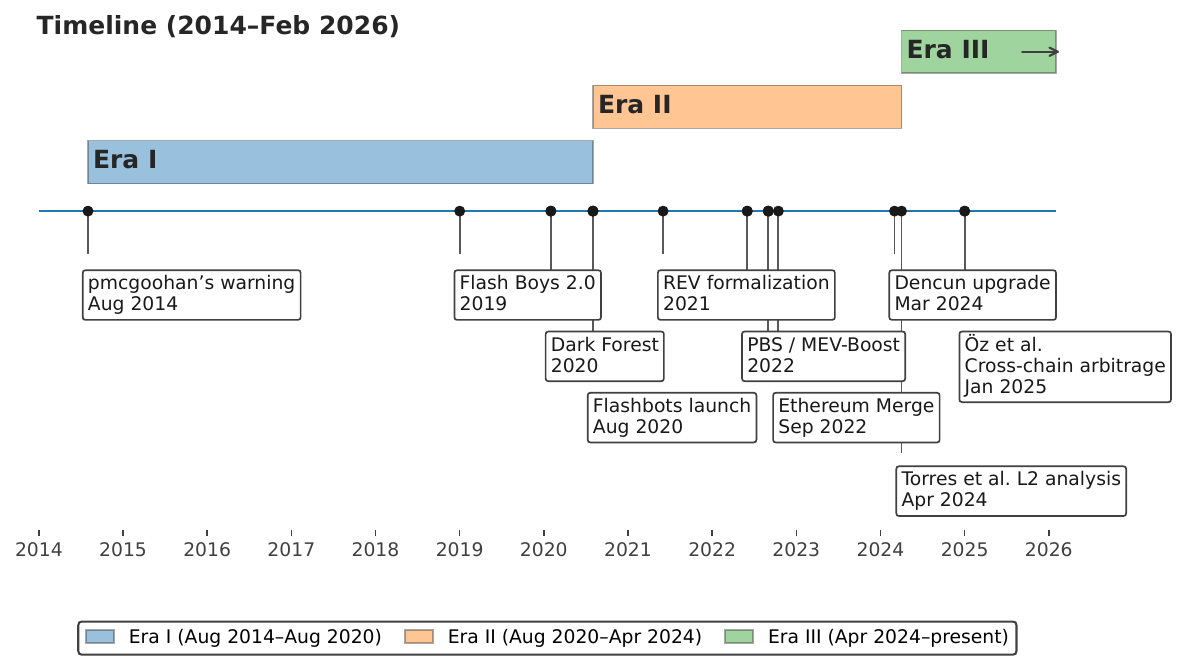}
\caption{Timeline of MEV evolution across three eras. The timeline illustrates the progression from Miner Extractable Value in Era~I through the industrialization of extraction in Era~II to the cross-chain frontier in Era~III, with key publications and infrastructure developments marked at their respective dates.}
\label{fig:timeline}
\end{figure}

This Systematization of Knowledge takes a historical perspective, explicitly tracking the evolution of MEV from its origins to its current cross-chain manifestations. Unlike surveys that organize literature by topic or technique, we structure our analysis around three chronological eras, each characterized by distinct conceptual shifts, empirical discoveries, and infrastructural responses. This approach reveals how academic understanding, industry practice, and protocol design have co-evolved in response to the challenges posed by extractable value. Figure~\ref{fig:timeline} provides a visual overview of this three-era framework, highlighting key milestones and the dominant paradigms that characterized each period.

The economic significance of MEV has grown substantially. Early measurements by Daian et al.~\cite{Daian2019Flash} identified over \$6 million in atomic arbitrages on Ethereum. Wu et al.~\cite{Wu2025CEXDEX} documented 233.8 million USD extracted through CEX-DEX arbitrage alone over a 19-month period (August 2023 -- March 2025). Cross-chain arbitrage has added another dimension, with Oz et al.~\cite{Oz2025CrossChain} identifying 242,535 cross-chain arbitrages generating 8.65 million USD in profit. These figures represent lower bounds, as many extraction strategies remain undetected or unmeasured.

\subsection{Motivation and Scope}

The MEV literature has grown rapidly but remains fragmented across multiple research communities, including cryptography, distributed systems, financial economics, and market microstructure. Terminology varies across papers, with different authors using distinct definitions for overlapping concepts. The transition from single-chain to cross-chain MEV introduces additional complexity, as value extraction now depends on coordinating actions across heterogeneous domains with different consensus mechanisms, latency characteristics, and trust assumptions.

Our systematization addresses three objectives. First, we consolidate scattered studies and terminologies into a unified taxonomy, clarifying what has been explored and identifying gaps. Second, we trace the causal connections between conceptual developments, showing how each era's challenges motivated the next era's innovations. Third, we provide practitioners and researchers with a roadmap for understanding the current state of cross-domain MEV and the open problems that remain.

The scope of this SoK encompasses academic publications, industry reports, and technical documentation spanning 2014 to 2025. We focus primarily on Ethereum and EVM-compatible chains, which represent the majority of MEV research, while noting relevant work on other blockchain architectures including Algorand~\cite{Oz2024FCFS}, Polygon~\cite{Vostrikov2025Polygon}, and various Layer-2 rollups~\cite{Bagourd2023L2MEV, Torres2024Rolling, Solmaz2025Optimistic}.

\subsection{Organizing Principles}

We organize the literature along two primary axes. The temporal axis spans from 2014 to 2025, divided into three eras: Era~I covers the birth and formalization of Miner Extractable Value (August 2014 -- August 2020), from pmcgoohan's Reddit warning through the ``Dark Forest'' article; Era~II addresses the generalization to Maximal Extractable Value (August 2020 -- April 2024), encompassing taxonomies, Realized Extractable Value, PBS, the Merge, and non-atomic MEV; and Era~III focuses on Cross-Chain MEV (April 2024 -- present), initiated by Torres et al.'s Layer-2 rollup analysis. The conceptual axis distinguishes between potential extractable value (what could be extracted given optimal ordering) and realized extractable value (what is actually extracted in practice), as well as between single-domain MEV (confined to one blockchain or ordering domain) and cross-domain MEV (requiring coordination across multiple domains).

This two-dimensional organization enables systematic positioning of research contributions. Works focused on theoretical bounds occupy the ``potential'' dimension, while empirical measurements populate the ``realized'' dimension. Similarly, single-chain analyses contrast with cross-chain studies that require correlating data across heterogeneous sources.

\subsection{Contributions}

This systematization makes several contributions to the MEV literature:

\begin{itemize}
\item We provide a comprehensive historical account of MEV's conceptual evolution, organized explicitly around three eras with clear transition points marked by seminal works. While prior surveys have cataloged MEV phenomena~\cite{Gramlich2024MEV, Yang2023SoK, McMenamin2023SoK}, our chronological framing highlights the causal connections between eras.
\item We synthesize definitions from across the literature, proposing a unified terminology that distinguishes key concepts including domains, sequencers, potential vs. realized MEV, and atomic vs. non-atomic extraction.
\item We present detailed taxonomies of MEV phenomena, techniques, and mitigations, mapped to the eras in which they emerged.
\item We compile empirical findings across studies, identifying methodological patterns and measurement challenges.
\item We outline open problems and a research agenda focused on cross-chain MEV detection, measurement standardization, and mitigation design.
\end{itemize}

\section{Preliminaries and Definitions}
\label{sec:prelim}

Before examining the historical evolution of MEV, we establish the foundational concepts and terminology that will be used throughout this systematization.

\subsection{Transaction Ordering and Block Production}

Blockchain systems execute state transitions through transactions that are grouped into blocks. The entity responsible for proposing a block, whether a miner in Proof-of-Work (PoW) or a validator in Proof-of-Stake (PoS), has discretion over which transactions to include and in what order. This ordering power is the fundamental source of extractable value, as certain transaction sequences yield profits that would not be available under alternative orderings.

In the simplest model, pending transactions reside in a public mempool visible to all network participants. Block producers select transactions based on fees, typically prioritizing higher-paying transactions. However, the visibility of pending transactions creates opportunities for strategic actors to observe profitable trades and insert their own transactions to capture value, a practice known as front-running~\cite{Eskandari2019SoK}.

The economics of block production have been extensively analyzed. Brolley and Zoican~\cite{Brolley2023OnDemand} compared high-frequency trading infrastructure costs between centralized and decentralized exchanges, showing that gas auction mechanics create distinct cost structures for speed acquisition. John et al.~\cite{John2025Economics} provided a comprehensive economic analysis of Ethereum post-Merge, examining staking rewards, EIP-1559 dynamics, and MEV distribution.

\subsection{Domains and Sequencers}

Following Obadia et al.~\cite{Obadia2021Unity}, we define a \textit{domain} as any system where a designated entity, called a \textit{sequencer}, controls the ordering of state-changing actions. A domain may be a Layer-1 blockchain, a Layer-2 rollup, a centralized exchange, a bridge protocol, or any other system with a defined ordering mechanism. The sequencer for a PoW chain is the miner who produces a block; for a PoS chain, it is the validator selected as proposer; for a rollup, it is typically a centralized or decentralized sequencer operator.

An \textit{action} in a domain is any operation that modifies state, such as a transaction, a swap, or a message relay. The set of reachable states from a given initial state depends on the sequence of actions executed. Extractable value arises when different action sequences lead to different final balances for the sequencer or collaborating parties.

Table~\ref{tab:domains} summarizes the key domain types and their sequencer characteristics. The diversity of sequencer implementations creates heterogeneous MEV landscapes across the blockchain ecosystem.

\begin{table*}[ht!]
\caption{Domain Types and Sequencer Characteristics}
\label{tab:domains}
\centering
\small
\begin{tabular}{p{2.5cm}p{3cm}p{3.2cm}p{2cm}}
\toprule
\textbf{Domain Type} & \textbf{Sequencer} & \textbf{Selection} & \textbf{Finality} \\
\midrule
L1 (PoW) & Miner & Computational race & Probabilistic \\
L1 (PoS) & Validator & Protocol selection & Deterministic \\
Optimistic Rollup & Sequencer operator & Centralized/Decentralized & Delayed \\
ZK Rollup & Prover/Sequencer & Centralized & ZK-verified \\
CEX & Exchange operator & Centralized & Instant \\
Bridge & Relayer/Validator set & Varies & Cross-chain \\
\bottomrule
\end{tabular}
\end{table*}

\subsection{Single-Domain Extractable Value}

Building on Babel et al.~\cite{Babel2023Clockwork}, single-domain extractable value refers to the profit available to a sequencer within a single ordering domain. Formally, given an initial state $s_0$ and a set of pending actions $A$, the single-domain MEV is the maximum profit achievable by selecting and ordering a subset of actions:
\begin{equation}
\text{MEV}_{\text{single}}(s_0, A) = \max_{\sigma \in \Sigma(A)} \text{profit}(s_0, \sigma)
\end{equation}
where $\Sigma(A)$ denotes all possible orderings of subsets of $A$. This definition captures potential extractable value, the theoretical maximum under optimal strategy.

Bartoletti et al.~\cite{Bartoletti2025Lean} provided the first mechanized formalization of MEV in the Lean theorem prover, enabling machine-checked proofs of upper bounds for extractable value. Their work includes the first machine-checked proof of optimality for sandwich attacks in Automated Market Makers.

\subsection{Cross-Domain Extractable Value}

Cross-domain MEV extends this concept to multiple domains~\cite{Obadia2021Unity}. When actions across domains $D_1, D_2, \ldots, D_n$ can be coordinated, extractable value may exceed the sum of single-domain values due to arbitrage opportunities, atomic execution guarantees, or information advantages spanning domains. The cross-domain MEV is defined as:
\begin{equation}
\text{MEV}_{\text{cross}}(s, A_1, \ldots, A_n) = \max_{\sigma_1, \ldots, \sigma_n} \sum_{i=1}^{n} \text{profit}_i(s_i, \sigma_i)
\end{equation}
where the maximization considers joint orderings across all domains, potentially requiring sequencer collusion or unified control.

The cross-domain formulation captures several important phenomena. When sequencers across domains can coordinate, they may extract value exceeding what any single sequencer could capture independently. This creates incentives for vertical integration and centralization across the multi-chain ecosystem.

\subsection{Potential vs. Realized Extractable Value}

A critical distinction in the literature is between potential MEV, the theoretical maximum given perfect information and execution, and realized MEV, the profit actually captured by searchers and block producers. The gap between potential and realized MEV arises from competition, execution failures, latency, and transaction costs. Empirical studies increasingly focus on realized MEV as the economically relevant quantity~\cite{Heimbach2024NonAtomic}.

Chi et al.~\cite{Chi2024Remeasure} remeasured MEV arbitrage and sandwich activity on Ethereum, arguing that prior methods under- or over-counted due to heuristic limitations. Their refined pipeline, designed to be robust in the post-Merge setting, estimates around \$675 million of pre-Merge MEV, highlighting how methodological choices affect measurement.

\subsection{MEV Primitives and Strategies}

MEV extraction can be understood through two fundamental \textit{primitives} that combine to form more complex strategies~\cite{Torres2021Frontrunner}. Understanding this compositional structure is essential for analyzing both existing and emerging MEV phenomena.

\subsubsection{Ordering Primitives}

The two basic ordering primitives are \textit{front-running} and \textit{back-running}. In \textit{front-running}, an attacker observes a pending victim transaction $T_V$ in the mempool and inserts their own transaction $T_A$ to execute \textit{before} $T_V$, profiting from the anticipated state change. In \textit{back-running}, the attacker's transaction $T_A$ is positioned to execute \textit{immediately after} $T_V$, capturing value from the state change induced by the victim's transaction. These primitives form the building blocks from which most MEV strategies are composed, though certain strategies such as time-bandit attacks rely on fundamentally different mechanisms (chain reorganization) that fall outside this compositional framework.

\subsubsection{Derived Strategies}

Complex MEV strategies emerge from combining these primitives. Table~\ref{tab:strategies} summarizes the major strategies and their compositional structure.

\begin{table*}[ht!]
\caption{MEV Strategy Types: Primitives and Composition}
\label{tab:strategies}
\centering
\small
\begin{tabular}{p{1.8cm}p{3cm}p{4.5cm}p{1.8cm}}
\toprule
\textbf{Strategy} & \textbf{Primitive(s)} & \textbf{Mechanism} & \textbf{Key Works} \\
\midrule
Arbitrage & Back-run & Exploit price differences post-trade & \cite{Wang2022Cyclic, McLaughlin2023Arbitrage} \\
Liquidation & Back-run & Seize collateral after price drop & \cite{Babel2023Clockwork} \\
Sandwich & Front-run + Back-run & Manipulate price around victim & \cite{Torres2021Frontrunner} \\
JIT Liquidity & Front-run + Back-run & Provide LP around large swap & \cite{DiNosse2025Stylized} \\
Time-bandit & Reorg & Re-mine blocks for historical MEV & \cite{Daian2019Flash} \\
CEX-DEX & Back-run (non-atomic) & Cross-venue price exploitation & \cite{Heimbach2024NonAtomic, Wu2025CEXDEX} \\
\bottomrule
\end{tabular}
\end{table*}

\textit{Arbitrage} is fundamentally a back-running strategy: the arbitrageur waits for a trade to create a price discrepancy across venues, then executes immediately after to capture the difference. Wang et al.~\cite{Wang2022Cyclic} formalized cyclic AMM arbitrage, identifying 292,606 cycles on Uniswap V2 with revenue exceeding 34,429 ETH. \textit{Liquidations} similarly employ back-running: the liquidator monitors collateral positions and executes immediately after prices move to trigger liquidation thresholds~\cite{Babel2023Clockwork}.

\textit{Sandwich attacks} represent the canonical combination of both primitives. The attacker front-runs the victim's swap with a transaction that moves the price unfavorably, then back-runs with a transaction that profits from the price movement induced by the victim. Torres et al.~\cite{Torres2021Frontrunner} empirically analyzed over 11 million Ethereum blocks, identifying \$18.41 million in profits from front-running attacks (displacement, insertion, and suppression).

\textit{Just-in-Time (JIT) liquidity} also combines both primitives but with a different objective: the JIT provider front-runs a large swap by adding concentrated liquidity, earns fees from the swap, and back-runs by removing the liquidity. Di Nosse et al.~\cite{DiNosse2025Stylized} identified statistical regularities in DEX trading associated with JIT providers and sandwich attackers, revealing distinctive market microstructure patterns on Uniswap v3.

\textit{Time-bandit attacks} represent a qualitatively different strategy involving blockchain reorganization. When MEV opportunities in past blocks exceed the cost of re-mining, miners may be incentivized to fork the chain and capture historical value~\cite{Daian2019Flash}. This links MEV directly to consensus security.

\section{Era I: Miner Extractable Value}
\label{sec:era1}

The first era of MEV research spans from August 2014 to August 2020, during which the foundational concepts were established in the context of Proof-of-Work Ethereum with a public mempool. This era begins with pmcgoohan's prescient warning and concludes with the ``Dark Forest'' article that brought MEV to mainstream attention.

\subsection{The Birth of MEV: pmcgoohan's Warning (August 2014)}

The MEV problem was first identified by the pseudonymous algorithmic trader \textit{pmcgoohan} in a Reddit post titled ``Miners Frontrunning'' on August 12, 2014~\cite{pmcgoohan2014}, remarkably \textit{one year before Ethereum's mainnet launch}. pmcgoohan warned that miners could observe pending transactions and insert their own transactions before and after a victim's trade to capture profit, describing what we now call a sandwich attack.

Vitalik Buterin responded directly to this post~\cite{pmcgoohan2014}, acknowledging the concern. This early warning established several key insights that would remain central to MEV research: (1) miners can reorder, delay, and insert transactions without controlling 50\% of hashpower; (2) the public mempool creates an information asymmetry favoring block producers; and (3) this is a \textit{fundamental} issue, not a bug to be patched.

\subsection{Pre-Formalization Observations}

Before the formal coining of ``MEV,'' researchers documented front-running as a concern in blockchain systems. Eskandari et al.~\cite{Eskandari2019SoK} provided an early systematization of front-running attacks, classifying them into three categories: displacement (replacing a victim transaction with an attacker's transaction), insertion (inserting transactions around a victim), and suppression (preventing transaction inclusion). This taxonomy established that unfair ordering was a practical concern across decentralized applications, though without a unified economic framework.

Front-running in blockchain systems mirrors analogous practices in traditional finance, where privileged actors exploit information or timing advantages~\cite{Gramlich2024MEV}. However, the transparent mempool of public blockchains created unprecedented visibility into pending transactions, democratizing access to front-running opportunities while simultaneously exposing users to exploitation. As Gramlich et al. noted, ``a similar and substantially richer variety of means to gain unethical profit from power asymmetries has appeared in the context of blockchain-based decentralized applications.''

The early DeFi ecosystem provided fertile ground for extraction. Eskandari et al.~\cite{Eskandari2019SoK} drew knowledge from the top 25 most active decentralized applications deployed on Ethereum, identifying front-running instances across DEXs, ICOs, and other applications. Their analysis of initial coin offerings showed evidence of abnormal miner behavior indicative of front-running token purchases.

\subsection{The Flash Boys 2.0 Moment}

The seminal work by Daian et al.~\cite{Daian2019Flash} introduced ``Miner Extractable Value'' as a formal concept. Through empirical analysis of Ethereum's mempool and on-chain data, they demonstrated the existence of Priority Gas Auctions (PGAs), competitive bidding wars among bots seeking to capture arbitrage opportunities. These auctions consumed block space, increased fees for regular users, and created negative externalities for the network.

Crucially, Daian et al. framed MEV as a consensus-level security concern. They introduced the \textit{time-bandit attack}, showing that sufficiently large MEV opportunities could incentivize miners to reorganize the blockchain and re-mine past blocks to capture historical value. This linked application-layer activity (DeFi trading) to protocol-layer security (consensus stability), establishing MEV as a systemic risk rather than merely an economic inefficiency.

The empirical measurements in Flash Boys 2.0 provided a conservative lower bound on MEV, identifying over \$6 million from ``pure-revenue'' atomic arbitrages. This figure, while substantial, represented only a fraction of total extractable value, as many strategies were not captured by their detection methodology.

\subsection{Negative Side Effects of MEV}

MEV extraction produces four categories of negative externalities that became increasingly apparent during Era~I~\cite{Torres2021Frontrunner}. \textit{Block congestion (spam)} results from the accumulation of failed bot transactions, as searchers submit redundant or speculative transactions that ultimately revert but still consume block space. \textit{High transaction fees} arise from priority gas auctions, where competing bots bid up gas prices to secure transaction ordering, pricing out regular users. \textit{Price manipulation} occurs through front-running and sandwich attacks that extract value directly from users' trades. \textit{Consensus instability} emerges when MEV opportunities become large enough to incentivize time-bandit attacks or other forms of chain reorganization.

Torres et al.~\cite{Torres2021Frontrunner} empirically documented these effects, finding evidence of systematic block stuffing where bots deliberately fill blocks to delay competitor transactions. Their analysis revealed that front-running activity was not merely opportunistic but involved sophisticated strategies to monopolize extraction opportunities.

\subsection{Characteristics of Era I}

Era~I is characterized by several distinctive features summarized in Table~\ref{tab:era1chars}. The consensus mechanism was Proof-of-Work, with miners as the central actors controlling transaction ordering. The mempool was public, providing symmetric information access to all participants willing to monitor it. MEV was primarily defined as potential value accessible to block producers, with limited measurement of realized extraction. The emphasis was on threats to consensus security, particularly the destabilizing effects of large MEV opportunities.

\begin{table}[!t]
\caption{Era I Characteristics}
\label{tab:era1chars}
\centering
\small
\begin{tabular}{p{3cm}p{4.8cm}}
\toprule
\textbf{Characteristic} & \textbf{Description} \\
\midrule
Consensus & Proof-of-Work \\
Central Actor & Miners \\
Mempool & Public, transparent \\
MEV Definition & Potential value to block producers \\
Primary Concern & Consensus security threats \\
Infrastructure & Primitive, gas-based competition \\
Measurement Focus & Lower bounds on atomic arbitrage \\
Key Strategies & Arbitrage, front-running, PGAs \\
\bottomrule
\end{tabular}
\end{table}

The infrastructure during this period was relatively primitive. Searchers competed through gas price bidding in the public mempool, leading to inefficient outcomes including failed transactions, wasted gas, and network congestion. The lack of private communication channels between searchers and miners meant that strategies were visible and easily copied.

\subsection{Early Mitigations}

Initial responses to MEV focused on application-layer defenses. Commit-reveal schemes aimed to hide transaction contents until ordering was finalized. Batch auctions proposed processing multiple orders simultaneously at uniform prices. However, these mitigations saw limited adoption during Era~I, as the infrastructure to support them was underdeveloped and the scale of MEV extraction was not yet fully appreciated.

\medskip
\noindent\textit{Bridge to Era~II}: On August 28, 2020, Dan Robinson and Georgios Konstantopoulos of Paradigm published ``Ethereum is a Dark Forest''~\cite{Robinson2020DarkForest}, bringing MEV to mainstream attention. This influential article introduced a powerful metaphor capturing the adversarial nature of Ethereum's mempool, marking the transition to Era~II and the generalization from ``Miner'' to ``Maximal'' Extractable Value.

\section{Era II: Maximal Extractable Value}
\label{sec:era2}

Era~II spans from August 2020 to April 2024 and is characterized by the generalization of MEV concepts, the development of formal taxonomies, theoretical models, and the industrialization of MEV extraction through Proposer-Builder Separation and sophisticated infrastructure. Figure~\ref{fig:pipeline} illustrates the dramatic transformation of the MEV extraction pipeline across these eras, from the simple public mempool model of Era~I to the complex supply chain that emerged in Era~II and the cross-chain coordination requirements of Era~III.

\subsection{``Ethereum is a Dark Forest'' (August 2020)}

The publication of ``Ethereum is a Dark Forest''~\cite{Robinson2020DarkForest} marked the beginning of Era~II. Robinson and Konstantopoulos recounted an attempt to rescue funds from a vulnerable smart contract, only to have the rescue transaction front-run by ``generalized front-runners'', i.e., bots that could detect and replicate any profitable transaction. This demonstrated that: (1) generalized front-runners can exploit \textit{any} profitable transaction, not just known patterns; (2) the mempool is an information-transparent environment where ``detection means certain death''; and (3) private transaction channels were needed for sensitive operations.

A follow-up post, ``Escaping the Dark Forest''~\cite{samczsun2020}, documented a successful \$9.6 million rescue operation using direct miner submission via SparkPool. These articles brought MEV to widespread attention beyond academic circles.

\subsection{From Miner to Maximal}

The transition from ``Miner Extractable Value'' to ``Maximal Extractable Value'' reflected the recognition that extraction opportunities were not specific to PoW mining. The renaming, which became the prevailing convention following the Flashbots research community's usage in 2021 (as surveyed by Gramlich et al.~\cite{Gramlich2024MEV}), captured a fundamental conceptual shift. In PoS systems, validators rather than miners control block production. In Layer-2 rollups, sequencers determine transaction ordering. The term ``maximal'' captured the protocol-agnostic nature of extractable value: any entity with ordering power could potentially capture it.

This renaming was more than semantic. It shifted the conceptual focus from a specific actor (the miner) to a structural property (the ability to extract value through ordering). This abstraction enabled analysis of MEV across diverse blockchain architectures and anticipated the multi-domain complexity that would emerge in Era~III.

\subsection{Taxonomies and Systematizations}

Era~II produced several comprehensive surveys and systematizations that organized the growing MEV literature. Gramlich et al.~\cite{Gramlich2024MEV} provided a systematic literature review consolidating divergent definitions and categorizing MEV manifestations. Their work synthesized and reviewed the existing state of knowledge, pointing to undiscovered areas relevant to the decentralized economy.

Yang et al.~\cite{Yang2023SoK} systematized MEV countermeasures, offering a taxonomy of over 30 mitigation approaches spanning encryption, sequencing, and economic mechanisms. Their framework organized countermeasures into four directions: MEV auction platforms, time-based order fairness, content-agnostic ordering, and MEV-aware application design.

McMenamin~\cite{McMenamin2023SoK} produced a systematization specifically focused on cross-domain MEV, examining protocols including shared sequencers, order-flow auctions, and batch settlements. This work identified multiple dimensions for categorizing MEV: how value is extracted, where it originates, who controls access, where extraction occurs, and how extracted value can be redistributed. McMenamin noted that ``there are many potential categorizations for MEV,'' including value-based, control-based, and redistribution-based taxonomies.

\subsection{Formalizations and Models}

Era~II also saw the development of formal theoretical frameworks for analyzing MEV. Babel et al.~\cite{Babel2023Clockwork} introduced automated analysis of economic security in smart contracts, enabling formal reasoning about extractable value in specific protocol contexts. Their Clockwork Finance framework provided tools for automated MEV analysis.

Obadia et al.~\cite{Obadia2021Unity} formalized cross-domain MEV, defining domains, sequencers, and the conditions under which cross-domain extraction exceeds single-domain extraction. Their key insight was that ``the MEV is the maximum of the sum of final balances across all considered domains into a single base asset, when some mix of actions across all those domains is executed together.''

Game-theoretic models examined the strategic interactions among searchers, block producers, and users. Capponi et al.~\cite{Capponi2025Allocative} showed that MEV induces inefficient block-space allocation; private pools can improve welfare but equilibrium adoption is partial because validators forgo rents. These models revealed that competitive MEV extraction could lead to inefficient equilibria, with resources wasted on competition rather than productive activity.

Schwarz-Schilling et al.~\cite{Schwarz2023Timing} modeled ``timing games'' in PoS, where proposers strategically delay block release within a slot to harvest time-dependent MEV while still gathering sufficient attestations. They estimated a marginal value of time of approximately 0.0065 ETH per second from relay bids.

\subsection{Realized Extractable Value (May 2021)}

In May 2021, Flashbots published ``Quantifying Realized Extractable Value''~\cite{Salles2021REV}, introducing the concept of \textit{Realized Extractable Value (REV)}. The key insight was that MEV is a theoretical quantity that can only be approached asymptotically, unforeseen extraction methods can and will be devised. Hence, REV represents the \textit{actual value extracted} from MEV opportunities, where REV $\leq$ MEV. This distinction became crucial for empirical measurement.

\subsection{Flashbots and Private Mempools}

The emergence of Flashbots in 2021 fundamentally transformed MEV infrastructure~\cite{Weintraub2022Flashbot}. Flashbots introduced private transaction submission channels that allowed searchers to bypass the public mempool, submitting bundles directly to miners (and later, builders). Within five months of deployment, mining pools representing over 97\% of Ethereum's hash rate had adopted Flashbots' mev-geth client, reaching 99.9\% after eight months.

Weintraub et al.~\cite{Weintraub2022Flashbot} empirically measured MEV extraction via Flashbots between blocks 10,000,000 and 14,444,725 (May 2020 -- March 2022). Their analysis revealed several key findings. First, \textit{profit redistribution} occurred: miners earned more using Flashbots while searchers earned less, as competition for bundle inclusion intensified. Second, \textit{extraction shifted to private channels}: 48\% of sandwich attacks were executed via Flashbots rather than the public mempool. Third, \textit{centralization emerged}: 90\% of Flashbots blocks were mined by just two mining pools, foreshadowing the builder concentration that would characterize the PBS era.

\subsection{Characteristics of Early Era II}

The early part of Era~II retained a focus on \textit{potential} extractable value rather than realized extraction. Research asked ``what could be extracted?'' rather than ``what is actually extracted?'' This emphasis was appropriate for establishing theoretical foundations but left empirical questions about actual extraction rates and actor behavior unresolved.

The period saw growing recognition that MEV was not uniformly harmful. Arbitrage contributed to price discovery and market efficiency. Liquidations maintained the solvency of lending protocols. The challenge was distinguishing value-creating from value-destroying MEV and designing mechanisms to encourage the former while mitigating the latter.

\subsection{The Ethereum Merge and MEV-Boost (September 2022)}

Ethereum's transition to Proof-of-Stake on September 15, 2022 (``The Merge'') fundamentally changed MEV dynamics. Block proposers became known in advance, changing game-theoretic considerations. MEV-Boost was activated 17 epochs after the Merge, implementing Proposer-Builder Separation.

\subsection{Proposer-Builder Separation}

\begin{figure}[!t]
\centering

\begin{subfigure}{\columnwidth}
  \centering
  \includegraphics[width=\columnwidth]{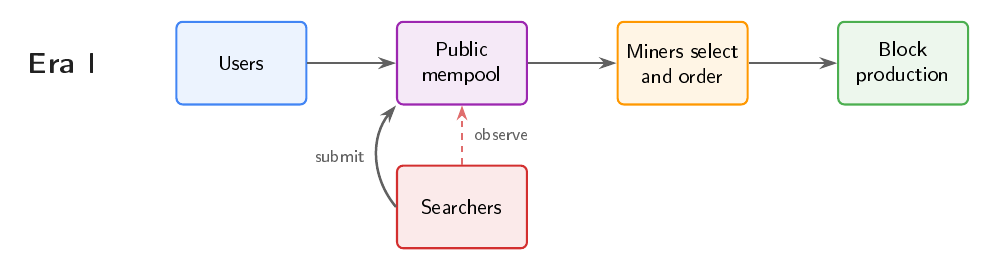}
  \caption{Era~I, public mempool and miner-controlled ordering.}
  \label{fig:pipeline-era1}
\end{subfigure}

\vspace{0.6em}

\begin{subfigure}{\columnwidth}
  \centering
  \includegraphics[width=\columnwidth]{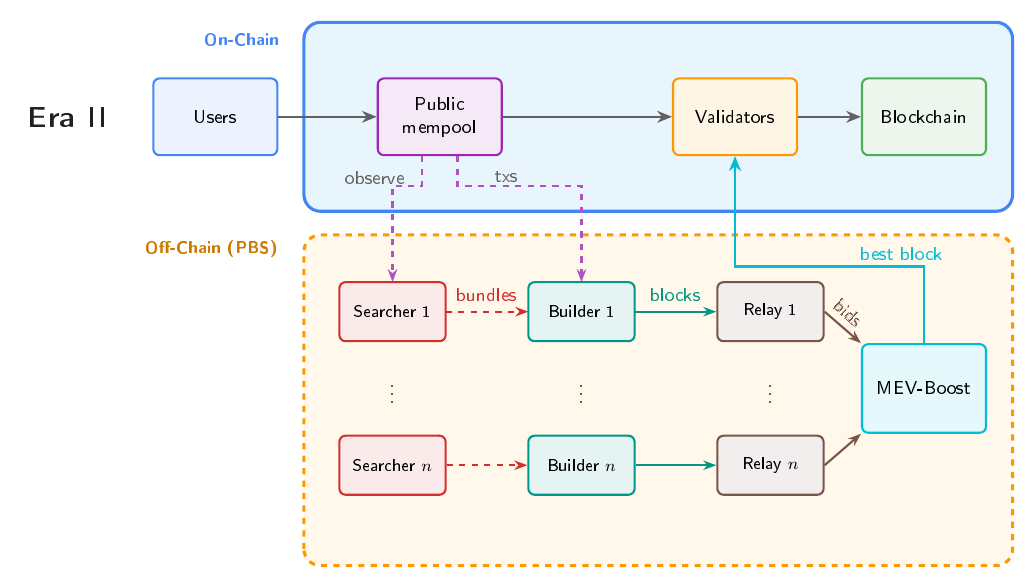}
  \caption{Era~II, PBS supply chain via builders, relays, and proposers (MEV-Boost).}
  \label{fig:pipeline-era2}
\end{subfigure}

\vspace{0.6em}

\begin{subfigure}{\columnwidth}
  \centering
  \includegraphics[width=\columnwidth]{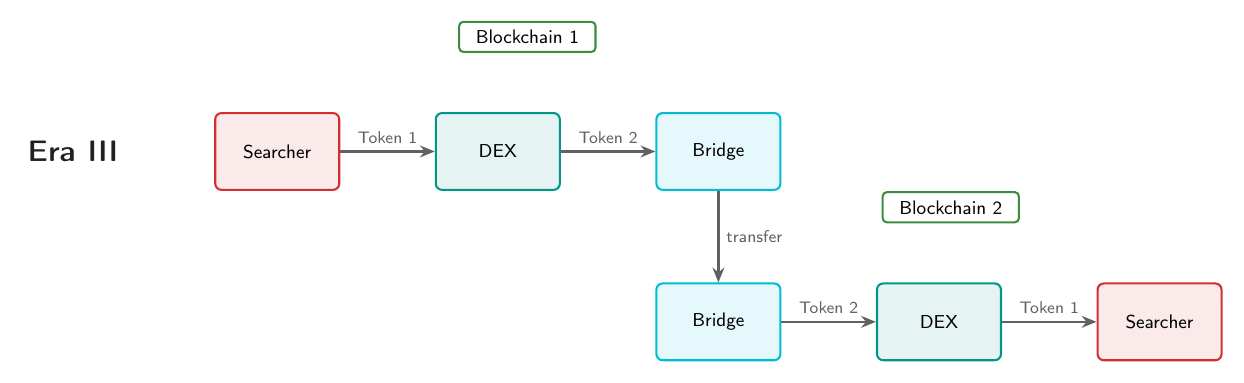}
  \caption{Era~III, cross-domain execution and settlement across multiple chains (SDA style).}
  \label{fig:pipeline-era3}
\end{subfigure}

\caption{Evolution of MEV extraction infrastructure. The figure compares the transaction flow from miner-dominated ordering in Era~I, to the PBS architecture in Era~II, and to the cross-chain of Era~III, where execution spans multiple chains.}
\label{fig:pipeline}
\end{figure}

The introduction of Proposer-Builder Separation (PBS) through MEV-Boost fundamentally restructured Ethereum's block production~\cite{Yang2025Decentralization, Heimbach2023PBS}. Under PBS, the roles of block building (selecting and ordering transactions) and block proposing (committing to a block) are separated. Specialized builders construct blocks to maximize value, and proposers select the highest-paying block through a competitive auction mediated by relays. As shown in Panel B of Figure~\ref{fig:pipeline}, this created a multi-stage supply chain involving searchers, builders, relays, and proposers.

PBS created a new supply chain for MEV. Searchers identify opportunities and submit bundles to builders. Builders aggregate bundles, add their own transactions, and construct complete blocks. Relays facilitate communication between builders and proposers while providing trust guarantees. Proposers select winning blocks and receive payments from builders.

Heimbach et al.~\cite{Heimbach2023PBS} measured PBS adoption and MEV dynamics in the post-Merge era. Within two months of the Merge, over 85\% of blocks were PBS blocks, with relay market share concentrated among a small number of operators. Their analysis of MEV profit distribution showed that validators engaging in PBS earned significantly more than non-participating validators, with proposers retaining approximately 90\% of profits compared to builders. Liquidation, sandwich, and arbitrage extraction overwhelmingly occurred through PBS blocks.

Empirical analysis revealed concerning concentration in this market. Yang et al.~\cite{Yang2025Decentralization} documented that two builders produced over 85\% of Ethereum blocks, creating centralization risks. Oz et al.~\cite{Oz2024WhoWins} analyzed builder auction dynamics from October 2023 to March 2024, finding that approximately 80\% of blocks were produced by three builders.

\subsection{Private Order Flow}

The proliferation of private transaction submission transformed MEV dynamics. Order Flow Auctions (OFAs) emerged as mechanisms for users to capture some of the MEV their transactions generate~\cite{Janicot2025Private}. Services like Flashbots Protect, MEV Blocker, Blink, and Merkle route transactions directly to builders, bypassing the public mempool.

Janicot and Vinyas~\cite{Janicot2025Private} noted that ``with the introduction of Proof of Stake and Proposer-Builder Separation, the transaction supply chain on Ethereum has shifted from relying entirely on the public mempool to an astonishing 80\% usage of private RPCs.''

Mancino et al.~\cite{Mancino2024Role, Mancino2025Favoritism} analyzed transaction ordering in the PBS era, uncovering systematic favoritism toward privately communicated transactions and hidden payments to builders.

\subsection{Non-Atomic and Off-Chain MEV}

Era~II extended MEV analysis beyond atomic, single-transaction strategies to non-atomic extraction spanning multiple transactions, blocks, or venues~\cite{Heimbach2024NonAtomic}. CEX-DEX arbitrage, exploiting price differences between centralized and decentralized exchanges, emerged as a major MEV source.

Heimbach et al.~\cite{Heimbach2024NonAtomic} measured non-atomic arbitrage on Ethereum DEXs from the Merge through October 2023, finding \$132 billion of volume representing nearly 30\% of DEX trading, with thousands of such trades occurring daily. Wu et al.~\cite{Wu2025CEXDEX} found 233.8 million USD extracted by 19 major searchers over 19 months.

\subsection{Layer-2 MEV Measurements}

The growth of Layer-2 rollups introduced new MEV dynamics. Bagourd and Francois~\cite{Bagourd2023L2MEV} quantified MEV on Polygon, Arbitrum, and Optimism, finding a lower bound of 213 million USD on Polygon alone.

Gogol et al.~\cite{Gogol2024CrossRollup} quantified non-atomic MEV opportunities across rollups, identifying over 500,000 unexploited arbitrage opportunities that persist on average for 10--20 blocks.

\medskip
\noindent\textit{Bridge to Era~III}: In April 2024, Torres et al.~\cite{Torres2024Rolling} published ``Rolling in the Shadows,'' the first systematic analysis of MEV across Layer-2 rollups, marking the transition to the cross-chain era. Their work demonstrated that MEV analysis must extend beyond single chains to encompass the multi-domain ecosystem.

\section{Era III: Cross-Chain MEV}
\label{sec:era3}

Era~III begins with Torres et al.'s ``Rolling in the Shadows'' (April 2024) and addresses MEV that arises from coordinating actions across multiple ordering domains. This represents the frontier of MEV research and the core focus of this systematization. Figure~\ref{fig:attacksurface} visualizes the complex attack surface that characterizes this era, depicting the interconnections between various domains and the MEV extraction paths that span them.

\subsection{Torres et al.: MEV Across Layer-2 Rollups (April 2024)}

Torres et al.~\cite{Torres2024Rolling} measured MEV across Ethereum and major rollups (Arbitrum, Optimism, zkSync) over approximately three years, finding that while MEV volumes were comparable to L1, profits and costs were lower, and no traditional sandwiching occurred due to the absence of public mempools on rollups. Crucially, they designed three cross-layer sandwich attacks using L1-emitted transactions targeting L2 trades, simulating approximately \$2 million potential profit. This work marks the formal beginning of Era~III by demonstrating that MEV extraction fundamentally depends on cross-domain coordination.

\subsection{Optimistic MEV on Layer-2}

Solmaz et al.~\cite{Solmaz2025Optimistic} introduced the concept of \textit{optimistic MEV}, i.e., MEV extraction through speculative transaction submission that exploits the unique properties of Layer-2 sequencers. Unlike traditional MEV where searchers observe pending transactions and react, optimistic MEV involves submitting transactions that \textit{anticipate} profitable opportunities will arise, accepting high failure rates in exchange for capturing value when opportunities materialize.

The empirical analysis revealed striking patterns in L2 blockspace consumption. On Base and Optimism in Q1 2025, optimistic MEV contracts, primarily executing cyclic arbitrage, consumed over 50\% of on-chain gas while paying less than 25\% of total transaction fees. This asymmetry arises because L2 sequencers typically employ first-come-first-served (FCFS) ordering with low base fees, allowing high-volume speculative submission at minimal cost per attempt.

The dominance of optimistic MEV explains why L2 blockspace demand remains persistently high despite low transaction fees. Searchers engage in continuous ``polling'' of the chain state, submitting arbitrage attempts that revert when no opportunity exists but succeed when price discrepancies arise. This creates a fundamentally different MEV landscape from L1: instead of competition through gas price bidding (PGAs), L2 MEV competition occurs through transaction volume and latency optimization.

\subsection{Foundations: Cross-Chain Communication}

Before defining cross-chain MEV, we must understand how chains communicate. Zamyatin et al.~\cite{Zamyatin2021SoK} provided the foundational systematization of cross-chain communication protocols, establishing that such communication is impossible without trusted third parties. All cross-chain designs involve trust assumptions, whether in federated validators, light clients, optimistic verification, or cryptographic proofs.

Deng et al.~\cite{Deng2025CrossChain} surveyed cross-chain interoperability, classifying designs into notary/relays, HTLC/hashed timelocks, light-client/IBC-style, and shared/sequenced rollups. They compared security, latency, and liveness trade-offs across architectures.

Bridges enable asset transfers between chains through various mechanisms. Notland et al.~\cite{Notland2025SoK} systematized bridge architectures and vulnerabilities based on 64 bridges, identifying 13 critical components and 8 vulnerability categories from 31 exploits and 4 known bugs. Belenkov et al.~\cite{Belenkov2025SoK} analyzed bridge hacks in 2022-2023, documenting attack patterns including custodian attacks and communicator attacks.

Li et al.~\cite{Li2025BridgeSecurity} surveyed cross-chain bridge security, cataloging vulnerabilities including smart-contract bugs, oracle manipulation, and centralization risks. They documented approximately \$2 billion stolen through bridge exploits across 13 incidents, the majority of which occurred in 2022. Zhao et al.~\cite{Zhao2023BridgeSecurity} provided a structured review decomposing architectures and trust layers, distilling four attack families from real incidents.

\subsection{Definitions of Cross-Domain MEV}

Obadia et al.~\cite{Obadia2021Unity} provided the first formal definition of cross-domain MEV, investigating whether extractable value depending on ordering across two or more domains jointly exists. They defined the MEV as the maximum of the sum of final balances across all considered domains, when some mix of actions across those domains is executed together.

\begin{figure}[ht!]
\centering
\includegraphics[width=0.9\columnwidth]{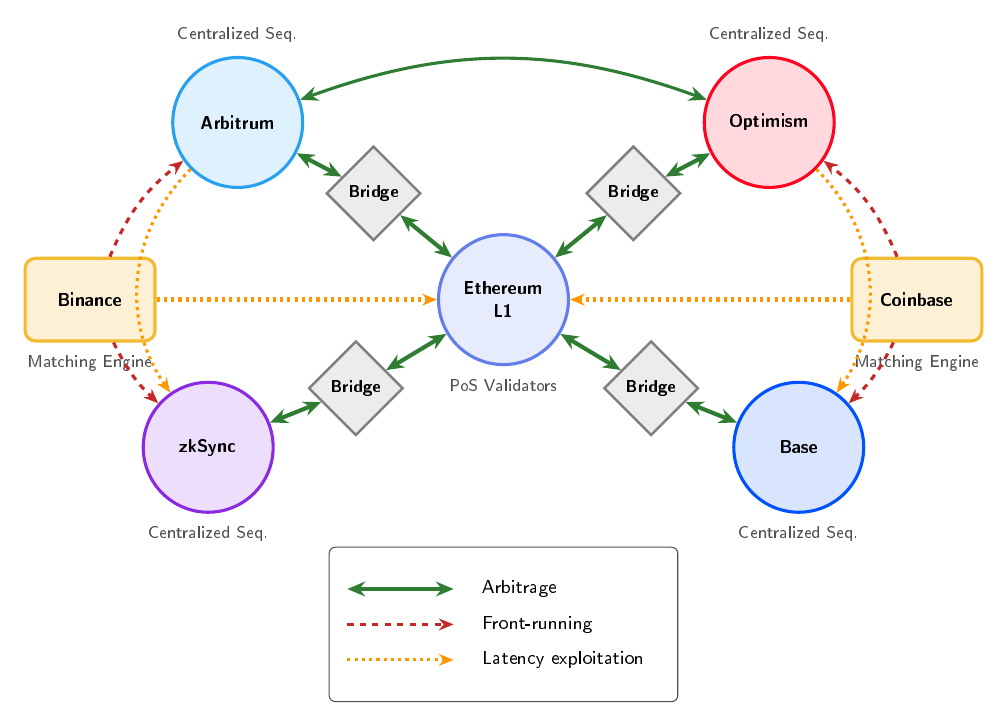}
\caption{Cross-Domain MEV Attack Surface. The graph depicts the multi-domain ecosystem where cross-chain MEV operates, with nodes representing ordering domains and edges representing value-extraction paths. Edge thickness reflects empirical extraction volume, while annotations indicate the types of MEV strategies applicable to each path. The visualization highlights how sequencer control at different nodes creates extraction opportunities spanning the ecosystem.}
\label{fig:attacksurface}
\end{figure}

The domain abstraction encompasses Layer-1 blockchains, Layer-2 rollups, centralized exchanges, bridge protocols, and sequencers. Cross-domain MEV arises when value extraction requires coordinated ordering across domains, not merely sequential execution within each domain. As illustrated in Figure~\ref{fig:attacksurface}, the edges connecting different domain nodes represent potential value-extraction paths, with annotations indicating the MEV types possible along each path.

McMenamin~\cite{McMenamin2023SoK} extended this framework by classifying cross-domain protocols and the MEV types each enables. The protocols analyzed include shared sequencers, multiple single-domain sequencers, order-flow auctions, app-chains, priority auctions, slot auctions, batch auctions, encrypted mempools, reduced block times, bridges, and off-chain RFQ systems.

Obadia et al.~\cite{Obadia2021Unity} identified several negative externalities of cross-domain MEV: cross-domain sequencer centralization, cross-domain time-bandit attacks, and the emergence of super-traders with systematic advantages.

\subsection{Empirical Measurement}

Empirical measurement of cross-chain MEV presents significant challenges. Unlike single-chain analysis, cross-chain detection requires correlating transactions across heterogeneous data sources with different schemas, timestamps, and finality guarantees.

Mazor and Rottenstreich~\cite{Mazor2024CrossChain} presented a framework for detecting cross-chain arbitrage between DEXs, analyzing PancakeSwap (BNB Chain) and QuickSwap (Polygon) over one month. They reconstructed pool states from factory events, modeled CPMM pricing, and searched cyclic paths. They found numerous unexploited opportunities with cross-chain revenues often exceeding single-DEX opportunities.

Sjursen et al.~\cite{Sjursen2023CrossDomain} moved toward quantifying cross-domain MEV, building a data-extraction tool spanning multiple EVM domains and identifying contracts actively engaging in cross-domain arbitrage.

Oz et al.~\cite{Oz2025CrossChain} conducted the first large-scale study of executed cross-chain arbitrage, introducing Sequence-Independent Arbitrage (SIA), where the arbitrageur holds pre-positioned capital on multiple chains, and Sequence-Dependent Arbitrage (SDA), where assets are moved across chains during execution. Analyzing nine blockchains over one year (September 2023 to August 2024), they identified 242,535 cross-chain arbitrages totaling 868.64 million USD in volume, generating 10.05 million USD in revenue and 8.65 million USD in net profit. Activity clustered on Ethereum-centric L1-L2 pairs, grew by 5.5$\times$ over the study period, and surged after the Dencun upgrade (March 13, 2024). Most trades used pre-positioned inventory (66.96\%) and settled in approximately 9 seconds, whereas bridge-based arbitrages (33.04\%) took approximately 242 seconds. Market concentration was high, with the five largest addresses executing over half of all trades.

\subsection{Sequence-Independent vs. Sequence-Dependent Arbitrage}

A key distinction in cross-chain MEV is between sequence-independent arbitrage (SIA), where the order of execution across chains does not affect profitability, and sequence-dependent arbitrage (SDA), where specific execution sequences are required~\cite{Oz2025CrossChain}. SIA opportunities can be captured without coordinated cross-chain ordering, while SDA requires precise timing and may involve bridge transactions that introduce latency and risk.

While Oz et al.~\cite{Oz2025CrossChain} focused on two-hop arbitrages involving pairs of blockchains, Mancino et al.~\cite{Mancino2025Multihop} extended the analysis to multihop (n-hop) cross-chain arbitrage, where execution spans three or more blockchains and is inherently sequence-dependent (SDA). Analyzing 12 blockchain networks and 45 bridges over a dataset of more than 2.4 billion transactions, their detection algorithm identified only 10 instances (8 three-hop, 2 four-hop) with no five-hop or six-hop arbitrages detected. Execution times averaged 434.1 seconds (7.2 minutes) for three-hop and 696.5 seconds (11.6 minutes) for four-hop arbitrages, with only 60\% profitability, underscoring the additional latency and coordination costs that each extra hop introduces.

\subsection{Negative Externalities}

Cross-domain MEV introduces several negative externalities~\cite{Obadia2021Unity}. \textit{Cross-domain sequencer centralization} occurs when maximizing cross-domain MEV requires controlling sequencers across multiple domains, creating pressure toward monopolistic or oligopolistic control. \textit{Cross-domain time-bandit attacks} extend the single-chain time-bandit concept, potentially incentivizing coordinated reorganizations across chains. \textit{Super-traders} with cross-domain execution capabilities gain systematic advantages over ordinary users.

The concentration of cross-chain arbitrage activity reinforces these concerns. Oz et al.~\cite{Oz2025CrossChain} found that five addresses executed over 50\% of trades, with one capturing approximately 40\% of daily volume post-Dencun. This concentration suggests barriers to entry and potential for coordinated behavior.

Landers and Marsh~\cite{Landers2025MCP} analyzed MEV in Multiple Concurrent Proposer (MCP) blockchains, formalizing novel MEV channels including same-tick duplicate steals, proposer-to-proposer auctions, and timing races. They noted that ``MCP chains move MEV from a single builder's private mempool to same-tick inter-block races created by data availability before final ordering.''

\subsection{Speculative MEV and Network Spam}

The absence of public mempools on rollups fundamentally changes MEV dynamics, giving rise to \textit{speculative MEV}, a qualitatively different extraction paradigm~\cite{Torres2024Rolling, Solmaz2025Optimistic}. In traditional (non-speculative) MEV, searchers monitor the public mempool off-chain, identify profitable opportunities, and submit targeted transactions. In speculative MEV (also termed \textit{optimistic MEV} by Solmaz et al.~\cite{Solmaz2025Optimistic}), the searching process moves on-chain: bots submit transactions speculatively, hoping to capture opportunities as they arise, with many transactions reverting when no opportunity exists.

This distinction has significant implications for network efficiency. Speculative MEV generates substantial \textit{spam}, i.e., transactions that do not change the final state of a block when removed. A transaction qualifies as spam if removing it from a block does not alter the final state in terms of balance and storage. Torres et al.~\cite{Torres2024Rolling} found that Optimism exhibits significantly higher levels of speculative arbitrage activity compared to Ethereum, with a larger number of both reverted and non-reverted arbitrage transactions. On average, speculative arbitrages proved more profitable but also contributed to significant block congestion.

The prevalence of speculative MEV on rollups reflects their architectural differences from Layer-1. With centralized sequencers operating on first-come-first-served (FCFS) ordering and no public mempool for transaction observation, searchers cannot employ traditional front-running strategies. Instead, they resort to continuous on-chain probing, creating a trade-off between extraction efficiency and network spam. As documented by Solmaz et al.~\cite{Solmaz2025Optimistic}, this behavior explains the persistent high demand for L2 blockspace: MEV bots on Ethereum's major private mempools operate predominantly through such speculative strategies on L2s, where low fees make high-volume polling economically viable.

\subsection{Cross-Layer Sandwich Attacks}

While traditional sandwiching is absent on rollups due to private sequencer mempools, Torres et al.~\cite{Torres2024Rolling} identified novel \textit{cross-layer sandwich attacks} that exploit the interaction between Layer-1 and Layer-2. These attacks leverage transactions emitted from L1 to L2 via bridge smart contracts, which are publicly visible on L1 before reaching the L2 sequencer.

Three attack strategies were identified. \textit{Strategy S1 (Classical Cross-Layer Sandwiching)} involves the attacker observing an L1-to-L2 swap transaction in the L1 mempool and submitting front-run and back-run transactions through the same bridge, sandwiching the victim on L2. \textit{Strategy S2 (Hybrid Sandwiching)} combines L1 bridge transactions for the front-run with direct L2 submission for the back-run, exploiting the latency between L1 emission and L2 inclusion. \textit{Strategy S3 (Speculative Sandwiching)} uses the bridge for the front-run but speculatively submits back-run transactions on L2, accepting potential reverts in exchange for faster execution.

Empirical validation on testnets confirmed the feasibility of all three strategies across Arbitrum, Optimism, and zkSync. Mainnet simulations estimated nearly \$2 million USD in potential profit, with Strategy S3 proving more effective for attackers with lower capital requirements. Transaction inclusion delays, averaging 798 seconds on Arbitrum Nitro and 83 seconds on Optimism post-Bedrock, create the temporal windows necessary for these attacks.

\subsection{Cross-Chain Mitigations}

Era~III has seen emerging infrastructure designed to address cross-chain MEV. Shared sequencer designs aim to provide atomic execution guarantees across rollups, potentially eliminating some cross-domain MEV while creating new forms~\cite{Silva2025Atomic}. Han et al.~\cite{Han2025SharedSequencer} proposed a shared sequencer model with multi-slot weighted leader election, PBS-style separation, and Fuzzy Cognitive Maps for parameter tuning.

Order-flow auctions extended to cross-chain contexts could redistribute extraction to users. Intent-based protocols abstract from specific execution paths, potentially reducing susceptibility to MEV. Shi et al.~\cite{Shi2024CoW} developed optimization models for intent settlement combining Coincidence of Wants with cross-chain AMM execution, reporting 26--46\% higher completion rates than CoW-only baselines.

Belchior et al.~\cite{Belchior2025Harmonia} proposed Harmonia, a ZK-based framework for secure cross-chain applications including a decentralized light client and ZK attestations. Belchior et al.~\cite{Belchior2024Hephaestus} introduced Hephaestus, a cross-chain model generator for monitoring applications and identifying malicious behavior.

However, as Silva and Livshits~\cite{Silva2025Atomic} demonstrated, atomic execution in shared sequencers does not necessarily eliminate arbitrage profits and can even induce losses depending on price paths and fees. The mitigation landscape remains immature, with theory and measurement racing to keep pace with evolving extraction strategies.

\subsection{Conceptual Closure}

Era~III completes the conceptual evolution of MEV. ``Miner Extractable Value'' denoted potential, single-chain, miner-centric extraction. ``Maximal Extractable Value'' generalized to any ordering entity while remaining largely focused on potential value. ``Realized Extractable Value'' captured actual extraction in practice. ``Cross-Chain MEV'' represents the subset of realized extractable value that fundamentally depends on coordinating across multiple domains. We propose \textit{Realized Cross-Chain Extractable Value} (RC-MEV) as the precise term for empirically measured cross-domain extraction.

\section{Timeline and Conceptual Taxonomy}
\label{sec:taxonomy}

This section provides structured representations of the MEV literature organized by temporal evolution and conceptual dimensions.

\subsection{Taxonomy of MEV Strategies}

Table~\ref{tab:taxonomy} classifies MEV strategies along three dimensions: the ordering primitives exploited (front-running, back-running, or both), whether execution is atomic (settled in a single transaction or block) or non-atomic, and the domain scope (single-chain or cross-chain).

\begin{table*}[ht!]
\caption{Taxonomy of MEV Strategies}
\label{tab:taxonomy}
\centering
\small
\begin{tabular}{p{3.615cm}p{1.18cm}p{1.515cm}p{1.3cm}p{7.2cm}}
\toprule
\textbf{Strategy} & \textbf{Primitive} & \textbf{Atomicity} & 
\textbf{Scope} & \textbf{Description} \\
\midrule
DEX-DEX arbitrage & Back-run & Atomic & Single & 
Cyclic swaps across pools in one 
transaction~\cite{Wang2022Cyclic} \\
CEX-DEX arbitrage & Back-run & Non-atomic & Cross & 
Price alignment between off-chain and on-chain 
venues~\cite{Heimbach2024NonAtomic, Wu2025CEXDEX} \\
Cross-chain arbitrage (SIA) & Back-run & Non-atomic & Cross & 
Pre-positioned inventory on multiple 
chains~\cite{Oz2025CrossChain} \\
Cross-chain arbitrage (SDA) & Back-run & Non-atomic & Cross & 
Bridge-mediated sequential 
execution~\cite{Oz2025CrossChain, Mancino2025Multihop} \\
Sandwich attack & Both & Atomic & Single & 
Front-run + back-run wrapping a victim 
swap~\cite{Torres2021Frontrunner} \\
Cross-layer sandwich & Both & Non-atomic & Cross & 
Speculative sandwich exploiting L2 sequencer 
delays~\cite{Torres2024Rolling} \\
Liquidation & Back-run & Atomic & Single & 
Collateral seizure triggered by price 
movement~\cite{Babel2023Clockwork} \\
JIT liquidity & Both & Atomic & Single & 
Concentrated LP added before and removed after a target 
swap~\cite{DiNosse2025Stylized} \\
Time-bandit & -- & Non-atomic & Single/Cross & 
Chain reorganization to capture historical 
MEV~\cite{Daian2019Flash} \\
\bottomrule
\end{tabular}
\end{table*}

\subsection{Chronological Overview}

Table~\ref{tab:timeline} presents a chronological overview of key developments across the three eras. For each period, we identify the conceptual shift, representative works, and infrastructure changes that characterized the era. This table complements Figure~\ref{fig:timeline} by providing detailed citations and descriptions for each era's defining characteristics.

\begin{table*}[ht!]
\caption{Timeline of MEV Evolution}
\label{tab:timeline}
\centering
\small
\begin{tabular}{p{2.2cm}p{3cm}p{5cm}p{5cm}}
\toprule
\textbf{Period} & \textbf{Conceptual Shift} & \textbf{Representative Works} & \textbf{Infrastructure Changes} \\
\midrule
Aug 2014 -- Aug 2020 (Era I) & Miner Extractable Value; MEV as consensus threat & pmcgoohan~\cite{pmcgoohan2014}, Eskandari et al.~\cite{Eskandari2019SoK}, Daian et al.~\cite{Daian2019Flash} & Public mempool, PGAs, early arbitrage bots, gas-based competition \\
\midrule
Aug 2020 -- Apr 2024 (Era II) & Maximal Extractable Value; REV; PBS industrialization & Robinson \& Konstantopoulos~\cite{Robinson2020DarkForest}, Salles~\cite{Salles2021REV}, Obadia et al.~\cite{Obadia2021Unity}, Yang et al.~\cite{Yang2025Decentralization}, Wu et al.~\cite{Wu2025CEXDEX} & Flashbots, MEV-Boost, relays, private order flow, OFAs, L2 sequencers \\
\midrule
Apr 2024 -- present (Era III) & Cross-Chain MEV; multi-domain coordination & Torres et al.~\cite{Torres2024Rolling}, Oz et al.~\cite{Oz2025CrossChain}, Mancino et al.~\cite{Mancino2025Multihop} & Bridges, shared sequencers, intent protocols, ZK proofs \\
\bottomrule
\end{tabular}
\end{table*}

\subsection{Conceptual Matrix}

Table~\ref{tab:matrix} organizes the literature along two dimensions: potential versus realized extractable value, and single-domain versus cross-domain scope. Each cell contains representative citations and brief descriptors.

\begin{table}[ht!]
\caption{Conceptual Matrix: Potential/Realized $\times$ Single/Cross-Domain}
\label{tab:matrix}
\centering
\small
\begin{tabular}{p{1.5cm}p{2.8cm}p{2.8cm}}
\toprule
& \textbf{Single-Domain} & \textbf{Cross-Domain} \\
\midrule
\textbf{Potential} & Theoretical bounds, optimal strategies, formal verification \cite{Babel2023Clockwork, Daian2019Flash, Bartoletti2025Lean} & Cross-domain optima, collusion analysis, sequencer incentives \cite{Obadia2021Unity, McMenamin2023SoK, Landers2025MCP} \\
\midrule
\textbf{Realized} & Empirical extraction, builder profits, CEX-DEX measurement \cite{Heimbach2024NonAtomic, Yang2025Decentralization, Wu2025CEXDEX} & SIA/SDA detection, multihop paths, bridge arbitrage \cite{Oz2025CrossChain, Mancino2025Multihop, Muradli2024CrossChain} \\
\bottomrule
\end{tabular}
\end{table}

\subsection{Representative Key Works by Era}

Table~\ref{tab:litmap} summarizes representative works for each era, highlighting their key contributions and methodologies.

\begin{table*}[ht!]
\caption{Representative Key Works by Era}
\label{tab:litmap}
\centering
\small
\begin{tabular}{p{2.6cm}p{0.7cm}p{1.3cm}p{6cm}p{5cm}}
\toprule
\textbf{Work} & \textbf{Era} & \textbf{Type} & \textbf{Key Contribution} & \textbf{Methodology} \\
\midrule
Daian et al.~\cite{Daian2019Flash} & I & Empirical & MEV formalization, PGAs, time bandit & Mempool analysis, on chain data \\
Eskandari et al.~\cite{Eskandari2019SoK} & I & SoK & Front running taxonomy & DApp analysis, case studies \\
Torres et al.~\cite{Torres2021Frontrunner} & I & Empirical & Large scale front running measurement & 11M plus blocks, \$18.41M profit \\
\midrule
Obadia et al.~\cite{Obadia2021Unity} & II & Formal & Cross domain MEV definition, domain abstraction & Mathematical formalization \\
Gramlich et al.~\cite{Gramlich2024MEV} & II & Survey & MEV categorization, definitions, open problems & Systematic literature review \\
Yang et al.~\cite{Yang2023SoK} & II & SoK & Countermeasure taxonomy & 30 plus approaches analyzed \\
Yang et al.~\cite{Yang2025Decentralization} & II & Empirical & PBS builder market concentration & Market structure analysis \\
\midrule
Torres et al.~\cite{Torres2024Rolling} & III & Empirical & MEV across L2s, cross-layer attacks, speculative MEV & Multi rollup measurement, attack design and simulation \\
Oz et al.~\cite{Oz2025CrossChain} & III & Empirical & SIA and SDA cross-chain arbitrage at scale & 9 chains, 1 year panel, profit and latency analysis \\
Mancino et al.~\cite{Mancino2025Multihop} & III & Empirical & Multihop cross-chain arbitrage rarity and execution costs & 12 chains, 45 bridges, 2.4B tx, graph correlation heuristics \\
Solmaz et al.~\cite{Solmaz2025Optimistic} & III & Empirical & Optimistic, speculative MEV on L2, gas and spam dynamics & On chain contract level measurement on rollups \\
\bottomrule
\end{tabular}
\end{table*}

\section{Mitigations and Design Space}
\label{sec:mitigations}

Each era of MEV evolution has motivated corresponding defenses. We organize mitigations by the era in which the underlying mechanism class became salient, noting that several systems level instantiations appeared later as the ecosystem matured.

\subsection{Era I Mitigations}

Early mitigations primarily targeted content based exploitation and simple ordering advantages by reducing actionable information at submission time.

\textit{Commit-reveal} requires users to publish a commitment first, then reveal the transaction content later, reducing content based front running during the commitment phase. Breidenbach et al.~\cite{Breidenbach2018Hydra} introduce \textit{Submarine Commitments}, a commit-reveal style countermeasure that temporarily conceals transactions on blockchains. In practice, commit-reveal introduces an additional round and therefore latency, it can worsen UX, and it must handle timeouts and reveal censorship.

\subsection{Era II Mitigations}

Era~II expanded the design space with cryptographic, protocol, and application level mechanisms that aim to blind block producers, constrain ordering, or reshape incentive compatibility.

\textit{Encrypted mempools} use threshold or delay encryption to hide transaction contents from block producers until ordering is finalized~\cite{Kavousi2025BlindPerm,Fernando2025TrX,Chakrabarti2025ST3}. These approaches can reduce content based MEV, but introduce key management and decryption availability assumptions, and may impose latency or throughput overheads.

Kavousi et al.~\cite{Kavousi2025BlindPerm} propose BlindPerm, combining encrypted submission with randomized permutation to mitigate producer side MEV, reporting reductions in arbitrage and sandwich extractability in simulation on historical data.

Fernando et al.~\cite{Fernando2025TrX} present TrX, integrating encrypted mempools with a BFT style setting, emphasizing low added latency via batched threshold encryption.

Chakrabarti et al.~\cite{Chakrabarti2025ST3} propose Silent Threshold Traitor Tracing, adding public accountability to threshold encryption for mempool privacy.

\textit{Fair ordering protocols} aim to constrain ordering according to timing or fairness notions rather than pure fee priority. Achieving meaningful fairness guarantees remains challenging under network delays and adversarial message scheduling~\cite{Alipanahloo2024Survey}. Alipanahloo et al.~\cite{Alipanahloo2024Survey} survey MEV mitigations across Ethereum and L2s, including fairness oriented approaches, and highlight the assumptions required for enforceable fairness definitions.

\textit{Batch auctions and batch based exchanges} reduce within batch ordering advantages by collecting orders over an interval and settling them at a uniform clearing price. McMenamin et al.~\cite{McMenamin2022FairTraDEX} propose FairTraDEX, a frequent batch auction based DEX that targets extractable value, including detailed Solidity and pseudo code designs. Batch based approaches can reduce certain ordering attacks, but introduce market structure complexity, additional latency, and adoption constraints that depend on liquidity concentration and integration into wallets, aggregators, and routing infrastructure.

\subsection{Era II Mitigations, Post Merge}

With the post Merge transaction supply chain dominated by PBS style pipelines, mitigations also shifted toward market design and revenue routing.

\textit{PBS} separates block proposing from block building, limiting proposer side manipulation while shifting influence to builders and relays. Empirical analyses report substantial builder concentration and discuss its implications for decentralization~\cite{Yang2025Decentralization}.

\textit{Order flow auctions} redistribute value by auctioning back run rights or by controlling information disclosure for searchers and solvers~\cite{Janicot2025Private,Passerat2025DP}. Passerat Palmbach and Wadhwa~\cite{Passerat2025DP} propose differentially private aggregate hints, enabling users to reason about privacy loss when sharing information.

\textit{MEV aware application design} modifies DeFi protocols to reduce extractability or to share extracted value. Braga et al.~\cite{Braga2024MEVSharing} propose protocol level MEV sharing via a dynamically adjusted extraction rate. Batch trading AMM designs can be viewed as application level instantiations of batch settlement ideas, for example FM AMMs~\cite{Canidio2025FMAMM}. These approaches tend to be protocol specific and face integration and adoption barriers.

\subsection{Era III Mitigations, Cross-Chain}

Cross-chain MEV introduces asynchronous settlement, bridging latency, and fragmented liquidity, so defenses remain nascent.

\textit{Shared sequencer designs} attempt to coordinate ordering across rollups, enabling stronger cross rollup guarantees in principle~\cite{Han2025SharedSequencer}. Silva and Livshits~\cite{Silva2025Atomic} argue that atomicity alone does not eliminate arbitrage incentives.

\textit{Intent based protocols} abstract execution away from explicit transaction paths, allowing solvers to compete to satisfy user intents~\cite{Shi2024CoW}. This can reduce some path based exploitation, but introduces solver market dynamics and potential censorship risks.

\textit{ZK based bridges} aim to reduce trust assumptions via cryptographic verification, potentially limiting some cross domain failure modes at the cost of prover complexity and computational overhead~\cite{Belchior2025Harmonia}.

Watts et al.~\cite{Watts2025FailureCosts} propose failure cost mechanisms for permissionless order flow auctions, penalizing solvers whose bids fail to execute, with extensions relevant to cross domain intents. Agrawal and Ribeiro~\cite{Agrawal2025Timelock} introduce Timelock Shield, combining VDF based timelock encryption with delayed execution to counter MEV and censorship.

Table~\ref{tab:mitigations} summarizes mitigations by era and mechanism type.

\begin{table*}[t]
\caption{MEV Mitigations by Era and Mechanism}
\label{tab:mitigations}
\centering
\small
\begin{tabular}{p{0.35cm}p{2.55cm}p{6cm}p{3.5cm}p{3.7cm}}
\toprule
\textbf{Era} & \textbf{Mechanism} & \textbf{Description} & \textbf{Key Works} & \textbf{Limitations} \\
\midrule
I & Commit reveal & Hide tx contents until ordered & Breidenbach et al.~\cite{Breidenbach2018Hydra} & Latency, UX friction \\
\midrule
II & Encrypted mempool & Threshold or delay encryption & \cite{Kavousi2025BlindPerm,Fernando2025TrX,Chakrabarti2025ST3} & Key management, latency \\
& Fair ordering & Time based ordering constraints & Alipanahloo~\cite{Alipanahloo2024Survey} & Synchronization assumptions \\
& Batch auctions & Uniform clearing prices & McMenamin et al.~\cite{McMenamin2022FairTraDEX} & Complexity, adoption \\
& PBS & Separate building and proposing & Yang~\cite{Yang2025Decentralization} & Builder centralization \\
& OFAs & Auction backrun rights & Janicot~\cite{Janicot2025Private}, Passerat~\cite{Passerat2025DP} & Searcher concentration \\
& App design & MEV resistant protocol design & Braga~\cite{Braga2024MEVSharing}, Canidio~\cite{Canidio2025FMAMM} & Protocol specific \\
\midrule
III & Shared sequencer & Atomic cross rollup execution & Han~\cite{Han2025SharedSequencer}, Silva~\cite{Silva2025Atomic} & New centralization \\
& Intents & Solver mediated execution & Shi~\cite{Shi2024CoW} & Solver competition \\
& ZK bridges & Verifiable cross chain bridging & Belchior~\cite{Belchior2025Harmonia} & Computational cost \\
& Failure costs & Penalize failed bids & Watts~\cite{Watts2025FailureCosts} & Design complexity \\
& Timelock shield & Timelock encryption plus delayed execution & Agrawal~\cite{Agrawal2025Timelock} & Added delay \\
\bottomrule
\end{tabular}
\end{table*}

\subsection{Mitigation Trade offs}

All mitigations involve trade offs. Encryption adds key management and availability assumptions, fair ordering relies on synchronization and network assumptions, PBS shifts centralization pressure to builders, shared sequencing can introduce new coordination hubs, intents depend on competitive solver markets, and cross domain settings add latency and trust surface via bridges. The design space remains open, especially for cross domain contexts.

\section{Open Problems and Research Agenda}
\label{sec:openproblems}

Our systematization reveals several open problems and directions for future research.

\subsection{Standardized Metrics}

The field lacks standardized metrics for MEV measurement. Different studies use incompatible definitions, detection methodologies, and data sources, making comparison difficult. Chi et al.~\cite{Chi2024Remeasure} demonstrated how methodological choices significantly affect MEV estimates. We call for community effort to establish benchmark datasets, standardized detection algorithms, and agreed-upon metrics for potential and realized MEV, both single-domain and cross-domain.

\subsection{Cross-Domain Detection}

Detecting cross-chain MEV requires correlating transactions across heterogeneous data sources. Current approaches rely on heuristics that may miss complex strategies or produce false positives. Mancino et al.~\cite{Mancino2025Multihop} noted the extreme scarcity of multihop opportunities, finding only 10 instances from 2.4 billion transactions. Formal frameworks for cross-domain transaction correlation, possibly leveraging graph-based methods or machine learning, represent an important research direction.

Seoev et al.~\cite{Seoev2025Bidding} proposed reinforcement learning frameworks for MEV extraction, suggesting that advanced ML techniques may be applicable to detection as well as extraction.

\subsection{Sequencer Incentives}

As rollups proliferate, understanding sequencer incentives becomes critical. How do sequencer designs affect MEV extraction and redistribution? What are the centralization pressures from cross-domain MEV on shared sequencer architectures? Game-theoretic analysis of sequencer competition and collusion remains under-developed.

Landers and Marsh~\cite{Landers2025MCP} began addressing these questions for MCP blockchains, but comprehensive models for heterogeneous multi-chain environments are lacking.

\subsection{Welfare Analysis}

The welfare implications of cross-domain MEV are unclear. While arbitrage contributes to price alignment~\cite{John2024Pricing}, other extraction strategies may represent pure transfers or create negative externalities. Capponi et al.~\cite{Capponi2025Allocative} showed MEV induces inefficient block-space allocation. Rigorous welfare analysis distinguishing value-creating from value-destroying MEV in cross-chain contexts would inform protocol design.

Castillo Le\'on and Lehar~\cite{CastilloLeon2025DeFi} surveyed empirical DeFi literature, documenting arbitrage frictions and cross-market segmentation that affect welfare analysis.

\subsection{Bridge Security and MEV}

Bridge vulnerabilities have resulted in billions of dollars in losses~\cite{Li2025BridgeSecurity, Notland2025SoK, Belenkov2025SoK}. The interaction between bridge security and MEV extraction is under-explored. Do MEV incentives create pressure toward less secure bridge designs? Can MEV-aware bridge architecture improve security?

Yan et al.~\cite{Yan2025CrossChainTx} conducted large-scale measurement of cross-chain bridges, quantifying costs and documenting ledger inconsistencies, providing data for security-MEV interaction analysis.

\subsection{Regulatory Implications}

As Ji and Grimmelmann~\cite{Ji2025Regulatory} discuss, MEV raises regulatory questions about market manipulation, fairness, and disclosure. They proposed an ``intent-respecting'' test: liability arises when actors profit by routing or ordering contrary to user intent.

The European Securities and Markets Authority~\cite{ESMA2025MEV} synthesized MEV concepts and market impacts, noting that ``existing countermeasures are judged incomplete'' and calling for continued supervisory attention.

Cross-chain MEV complicates regulatory analysis by spanning multiple jurisdictions and potentially involving both decentralized and centralized venues.

\subsection{Network-Level Attacks}

Doumanidis and Apostolaki~\cite{Doumanidis2025Routing} studied routing attacks on Ethereum PoS, revealing strong network centralization with approximately 60\% of validators in just 100 IP prefixes. They proposed attacks yielding up to +71.4\% proposer rewards, highlighting infrastructure vulnerabilities that interact with MEV incentives.

Bayan et al.~\cite{Bayan2025Privacy} provided comprehensive analysis of privacy challenges in permissionless blockchains, examining threats including MEV extraction and signature vulnerabilities.

\section{Conclusion}
\label{sec:conclusion}

This Systematization of Knowledge has traced the evolution of Maximal Extractable Value through three distinct eras. Era~I (August 2014 -- August 2020) established MEV as an ordering problem in single PoW chains, from pmcgoohan's prescient warning through the ``Dark Forest'' recognition, identifying it as a threat to consensus security. Era~II (August 2020 -- April 2024) generalized MEV to any ordering entity, encompassing Realized Extractable Value, Proposer-Builder Separation, the Ethereum Merge, MEV-Boost, and non-atomic CEX-DEX arbitrage, documenting the industrialization of extraction. Era~III (April 2024 -- present), initiated by Torres et al.'s Layer-2 analysis, extended the field to cross-chain contexts, where value extraction depends on coordinating actions across multiple domains.

The relevant unit of analysis is no longer a single chain but the multi-domain ecosystem encompassing Layer-1 blockchains, Layer-2 rollups, centralized exchanges, bridges, and sequencers. Cross-domain MEV represents both a frontier for extraction strategies and a challenge for mitigation design. Theory and measurement are racing to catch up with an evolving landscape.

Our systematization provides researchers and practitioners with a unified framework for understanding MEV's past and present. The open problems we identify, spanning standardized metrics, cross-domain detection, sequencer incentives, welfare analysis, bridge security, and regulatory implications, constitute a research agenda for the field. As blockchain ecosystems continue to grow in complexity, understanding and managing extractable value will remain central to achieving the goals of decentralization, fairness, and security.

\bibliographystyle{IEEEtran}
\bibliography{refs}

\end{document}